\documentclass[
reprint,
groupedaddress,
amsmath,
amssymb,
aps,
prb,
longbibliography,
noeprint,
raggedbottom
]{revtex4-2}

\setcitestyle{super}

\usepackage{epsfig}
\usepackage{dcolumn}
\usepackage{bm}
\usepackage{color}
\usepackage{graphicx}
\usepackage{bbold}
\usepackage{lineno}

\begin{document}

\title{Microscopic scales and mechanism of quantum phase transitions in two-dimensional superconducting systems}

\author{Andrey Rogachev} 
\affiliation {Department of Physics and Astronomy, University of Utah, Salt Lake City 84093, USA}
\author{Samuel Feldman} 
\affiliation {Department of Physics and Astronomy, University of Utah, Salt Lake City 84093, USA}
\author{Kevin Davenport}
\affiliation {Department of Physics and Astronomy, University of Utah, Salt Lake City 84093, USA}

\date{\today}

\begin{abstract} 
\textbf{The superconducting ground state in many two-dimensional materials can be created or destroyed through quantum phase transitions (QPTs)\cite{SacepeKlapwijk,Goldman_PC_review,SondhiShahar_QPR_review} controlled by non-thermal parameters such as carrier density or magnetic field. While various mechanisms for these QPTs have been proposed,\cite{Gantmakher_SIT_review} it remains unclear which, if any, are applicable to a specific two-dimensional superconducting system.  Here, we find that a pair-breaking mechanism \cite{Tinkham_book,DelMaestro_AnnPhys} which suppresses the Cooper pair density gives a unifying description of magnetic-field-driven QPTs in amorphous MoGe, Pb and TaN films, and the high-temperature superconductor La$_{1.92}$Sr$_{0.08}$CuO$_{4}$. This transition occurs within the superconducting subsystem and is masked by the dominant non-critical contribution of normal electrons. The discovery was enabled by the development of a QPT model that goes beyond the conventional determination of the critical exponents and incorporates into the analysis a microscopic length scale characterizing the transitions. We found that in the materials studied, and MoGe nanowires, this scale corresponds to the size of a Cooper pair. The model has also been successfully applied to QPTs in Josephson junction arrays\cite{Rogachev_QPT_JJ} and various non-superconducting materials. \cite{Rogachev_MicroScale} The observation that microscopic scales are encoded in the scaled experimental data of QPTs likely extends beyond equilibrium condensed matter physics and may reveal underlying principles of critical phenomena in a wide variety of systems.\cite{Non-Eq_PT_book,Heyl_DynamicalQPT,Munoz_PTliving}}

\end{abstract}
\maketitle
\runningpagewiselinenumbers
Two-dimensional (2D) superconductors constitute a large, rapidly growing class of materials ranging from thin films of conventional and high-temperature superconductors \cite{Dynes, BaturinaStrunk, HollenValles, Szabo_MoC_STM, BollingerBozovic} to interface, surface, hybrid superconductors, and twisted bi-layer graphene. \cite{Caviglia LAOSTO, SaitoNojima, BottcherMarcus, CaoHerrero}  Understanding QPTs in these systems is a challenging task. \cite{SacepeKlapwijk} The corresponding many-body 2D problems pose significant challenges for theoretical understanding. The complexity of real 2D materials, which often  exhibit disorder and a tendency to form spatially non-uniform states, \cite{SacepeKlapwijk,Chand_Pratab_NbN_STM,Carbillet_Rodichev_STM_NbN,BoudimTrivedi} further complicate the field.

In proximity to a continuous QPT's critical point, the quantum system becomes scale-invariant. \cite{SondhiShahar_QPR_review} In apparent agreement with this prediction, generic scaling behavior of conductivity has been observed in many 2D superconductors. \cite{Hebard_Paalanen,Yazdani_Kapitulnik_SIT,MarkovicGoldman,OvadiaShahar} However, the critical exponents obtained from the scaling analysis fail to exhibit the expected universality. \cite{SacepeKlapwijk} Furthermore, scaling behavior alone does not disclose the microscopic mechanism of QPTs \cite{Goldman_PC_review} and, in some cases, may even be accidental. \cite{Rogachev_Deficiency}

There are several microscopic scenarios which can lead to a QPT in a 2D superconductor. The most well-known, and most often used for comparison with experiments, is the dirty boson model. \cite{Fisher_SIT} It proposes that the transition occurs due to the delocalization of superconducting vortices and implies that the order parameter amplitude, the quantity which represents the density of the Cooper pairs, remains constant across the QPT and the superconductivity is destroyed entirely by phase fluctuations.

\begin{figure*}[tbph]
\centering
 \includegraphics[width= 1.0\textwidth]{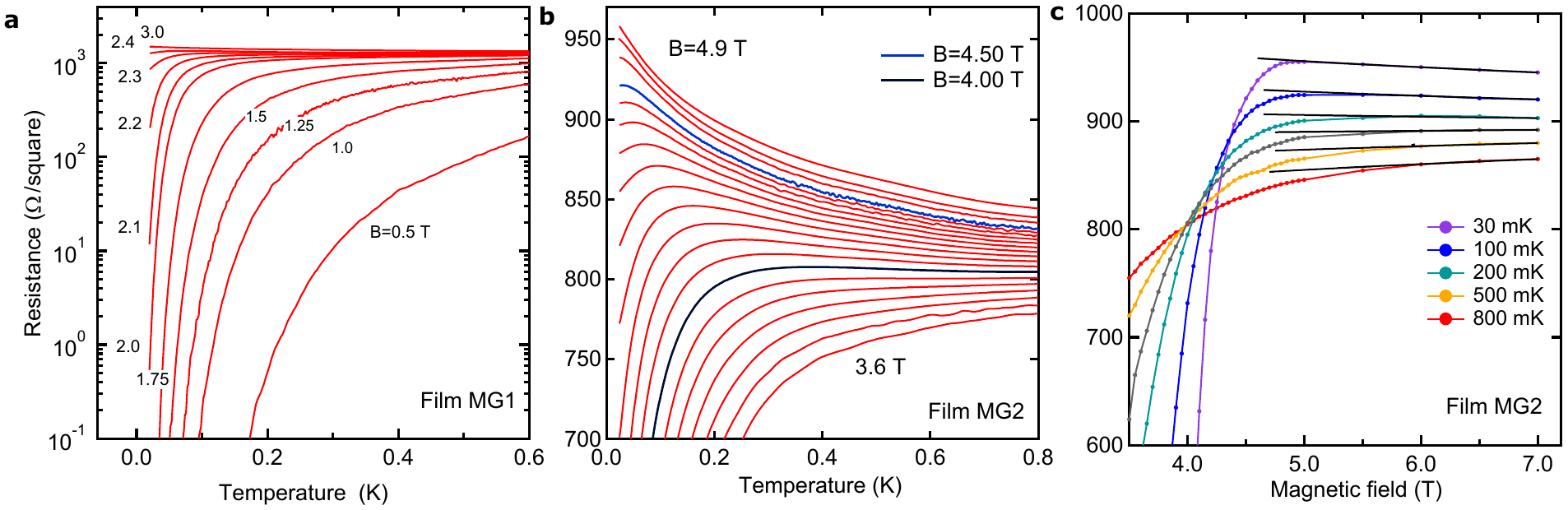}
 \caption{\textbf{Effect of magnetic field on superconducting MoGe films.} 
\textbf{(a,b)} Resistance versus temperature for film MG1 (composition Mo$_{50}$Ge$_{50}$, nominal thickness 2.5 nm, $T_{c}$=1.2 K) and MG2 (Mo$_{78}$Ge$_{22}$, 2.5 nm, $T_c$=2.8 K) . \textbf{(c)} Resistance versus magnetic field for MG2 film. The black solid lines are the linear fits to the data in the range 6-7 T extended to lower fields. They were used to estimate conductance of normal electrons.}
 \end{figure*}

We have recently discovered, however, that in superconducting $\it{nanowires}$ the magnetic-field-driven QPT can be described by an alternative pair-breaking critical theory. \cite{Rogachev_PBnanowire,DelMaestro_AnnPhys} Microscopically, the pair-breaking processes split Cooper pairs by acting on their orbital and spin components. Cooper pair density goes to zero at the critical field, $B_{c}$, and the critical fluctuations are of Aslamazov-Larkin type involving both amplitude and phase.

Guided by the expectation that the magnetic-field-induced pair-breaking processes are similar in 1D and 2D, \cite{Tinkham_book} we have studied and analyzed the transitional regime in several superconducting 2D systems and discovered pronounced pair-breaking QPTs in amorphous MoGe, Pb and TaN films as well as in the high-temperature superconductor La$_{1.92}$Sr$_{0.08}$CuO$_{4}$ (LSCO). 

\smallskip
\noindent \textbf{Pair-breaking quantum phase transition.}

The details of the fabrication and measurements of the films made of amorphous MoGe alloy are given in the supplementary information (SI). The data for the amorphous Pb films were traced from Ref. [\cite{Xiong_Book}], for amorphous TaN from Ref. [\cite{BreznayKapitulnik_TaN}], and for LSCO from Ref. [\cite{AndoKishio_LSCO_QPT}]. The magnetic field was oriented normal to the films and to CuO$_{2}$ layers in LSCO.

Figure 1(a) displays the temperature dependence of the resistance of MoGe film MG1. In the superconducting state, all curves display a roughly exponential variation with no leveling to a finite value.  This contrasts with the resistance saturation previously observed in MoGe films, as well as in other systems, which has been taken as evidence for an “anomalous metal” regime; see Ref. [\cite{KapitulnikSpivak_AnomalousMetal}] for a review. In our view, this saturation is controversial; in some systems, it represents a real physical process \cite{BottcherMarcus} but in the others, an artefact caused by insufficient noise filtering. \cite{Tamir_Shahar_NoStrangeMetal}

Let us now consider the transitional regime between the superconducting and metallic states.  The phenomenological finite-size scaling theory \cite{SondhiShahar_QPR_review} explains how the presence of a QPT at zero temperature affects the response of a system at finite temperatures.  Near a magnetic-field-driven QPT, the spatial correlation length of the quantum fluctuations, $\xi$, diverges as $\xi\propto\left|B-B_c\right|^{-\nu}$, where $\nu$ the correlation length critical exponent. The dynamics of the fluctuations are characterized by a temporal scale $\xi_\tau$, related to $\xi$ as  $\xi_{\tau}\propto\xi^z$, where $z$ is the dynamical critical exponent. The temperature cuts off the quantum fluctuation time scale and sets the dephasing length in the system $L_{\varphi}\sim T^{-1/z}$. The relation $L_\varphi\lesssim\xi$
 defines the boundary of the quantum critical regime, where the conductivity follows the equation:
\begin{equation}
\sigma(B,T)=\frac{e^2}{\hbar} L_{\varphi}^{-(d-2)} \Phi_{\sigma \pm}\left(\frac{|B-B_c|}{T^{1/z\nu}}\right).
\end{equation}
Here $d$ is dimensionality of the system and $\Phi_\sigma$ is the scaling function (unknown within this model).

In Figure 1(b), we plot resistance of film MG2 versus temperature across the transition; the data are replotted in Fig. 1(c) as a function of magnetic field. According to Eq. 1, for a 2D system ($d$=2), the prefactor is a constant and, at $B$=$B_c$, the conductivity is independent of temperature. Experimentally, this means that there must be a temperature-independent separatrix between the superconducting-like and insulating-like $R\left(T\right)$ curves in Fig.1(b) and a single-point crossing for the $R(B)$ curves in Fig.1(c). Neither behavior is, in fact, observed. The problem is rooted in the assumption that $\textit{all}$ film resistance comes from superconducting fluctuations. This assumption is natural for a “phase-only” QPT and is valid for Josephson junction arrays but is questionable for superconducting films. 

An alternative pair-breaking mechanism of QPTs requires that the density of Cooper pairs goes to zero at $B_c$ and superconductivity becomes gapless, coexisting with normal electrons. In our analysis, which implies the pair-breaking mechanism from the start, we adopt the two-fluid model and assume that the conductivity of the film, $\sigma_{exp}$, comes from both the normal, $\sigma_n$, and superconducting, $\sigma_{sc}$, channels. The estimate of $\sigma_n$ was obtained from a linear extrapolation of $R\left(B\right)$ curves from high to low fields as shown by the solid lines in Fig. 1(c).  The conductivity $\sigma_{sc}$ was then obtained by the simple subtraction, $\sigma_{sc}$=$\sigma_{exp}-\sigma_n$.  The same procedure was applied to TaN and Pb films; as an example, the analysis of the latter system is presented in details in SI. 
\begin{figure*}[tbph]
\centering
 \includegraphics[width= 1.0\textwidth]{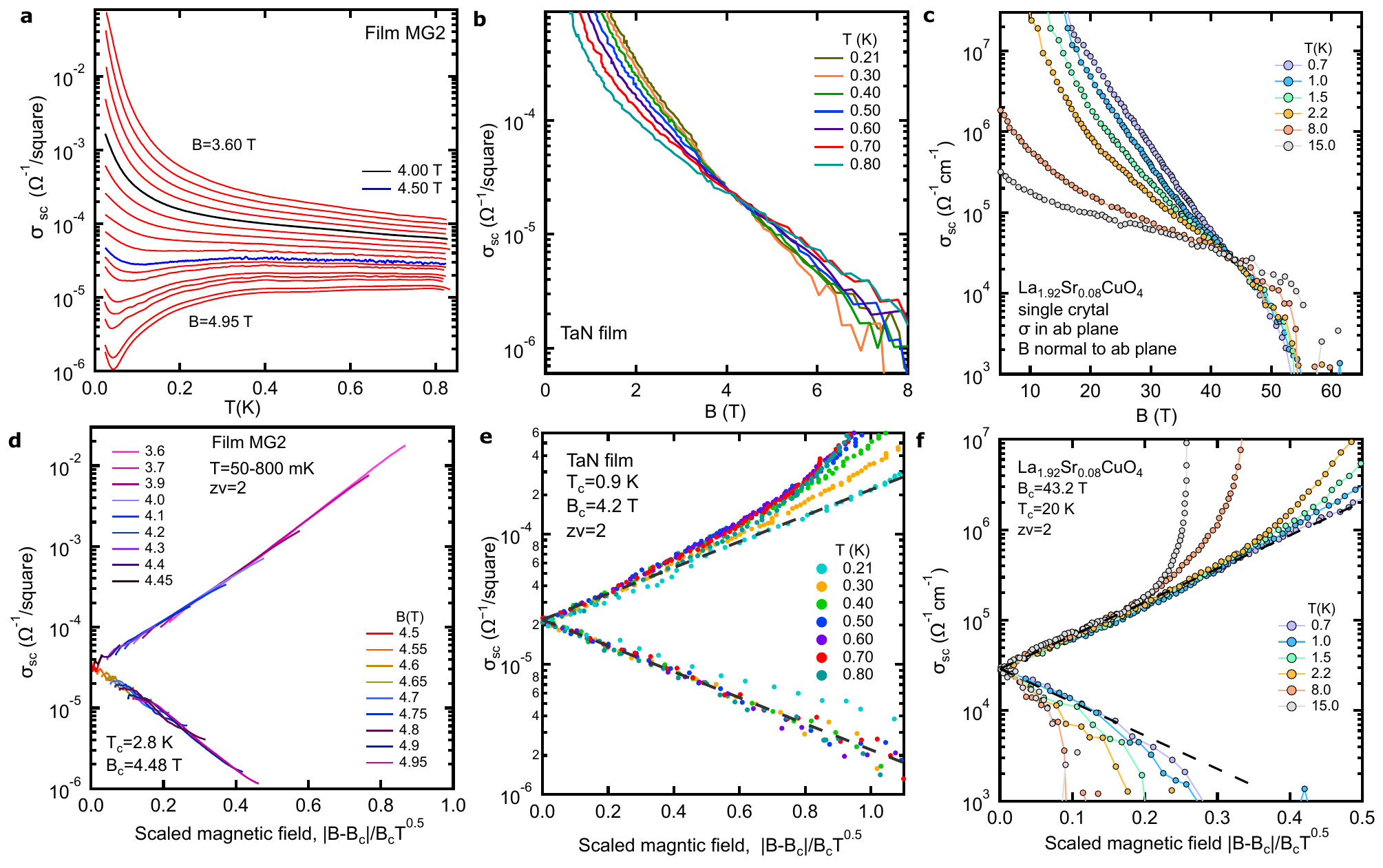}
 \caption{\textbf{Scaling in MoGe, TaN films and LSCO single crystal.} 
The conductivity of the superconducting channel versus temperature \textbf{(a)}, and magnetic field \textbf{(b,c)} for indicated samples. \textbf{(d-f)} The same conductivity versus scaled magnetic field. The dashed lines in \textbf{(e)} and \textbf{(f)} indicate the exponential “mirror symmetric” variation and have the same absolute slope.}
 \end{figure*}

We have also added to the analysis the magnetoresistance data obtained for a single crystal of La$_{1.92}$Sr$_{0.08}$CuO$_{4}$ (LSCO) measured in magnetic field up to 61 T. \cite{AndoKishio_LSCO_QPT} LSCO has a layered quasi-2D structure; in the first approximation, it can be considered as a system of semi-isolated two-dimensional superconducting  CuO$_{2}$ planes.  The original magnetoresistance data (Fig.1 in Ref. [\cite{AndoKishio_LSCO_QPT}]) resemble the data for MG2 film shown in Fig.1(c). For LSCO, $\sigma_n$ was estimated from linear extrapolation of  $R\left(B\right)$ curves between $B$=55 and 61 T. The data at 4.2 K appeared as an outlier in these fields and have not been included in the analysis. 

The conductivity of the superconducting channel of the MG2 film, $\sigma_{sc}$, is plotted versus temperature in Fig. 2(a). The striking feature of the graph is a clear presence of the insulating (low) branch.  The deviation from it, which appears as an upturn below about 30-50 mK, is likely an artefact of our subtraction procedures or is caused by the thermal decoupling of the electron subsystem. For TaN film, $\sigma_{sc}$ is plotted against magnetic field in Fig.2(b); the data reveals the expected crossing of $\sigma_{sc}\left(B\right)$ curves. A similar crossing, but at a much higher field, is observed in an LSCO single crystal (Fig.2(c)). 

According to Eq. 1, in the critical regime, all conductivity data should “collapse” onto a single curve when plotted against the scaled magnetic field. 

 The major result of this work is shown in Fig.2(d-f) which displays $\sigma_{sc}$ versus the scaling variable $|B-B_c|/B_c T^{1/z\nu}$. A collapse of the data is obtained for the $\it{same}$ value of  $z\nu\approx2$ for three MoGe, two Pb, one TaN films and surprisingly also for the LSCO single crystal. For MoGe films and LSCO, the scaling collapse occurs in an impressive temperature range exceeding one order of magnitude. Moreover, for MoGe films, where we have high-quality resistance data extended to very low temperatures, it covers 3-4 orders of magnitude of conductivity. 

\begin{figure*}[tbph]
\centering
 \includegraphics[width= 1.0\textwidth]{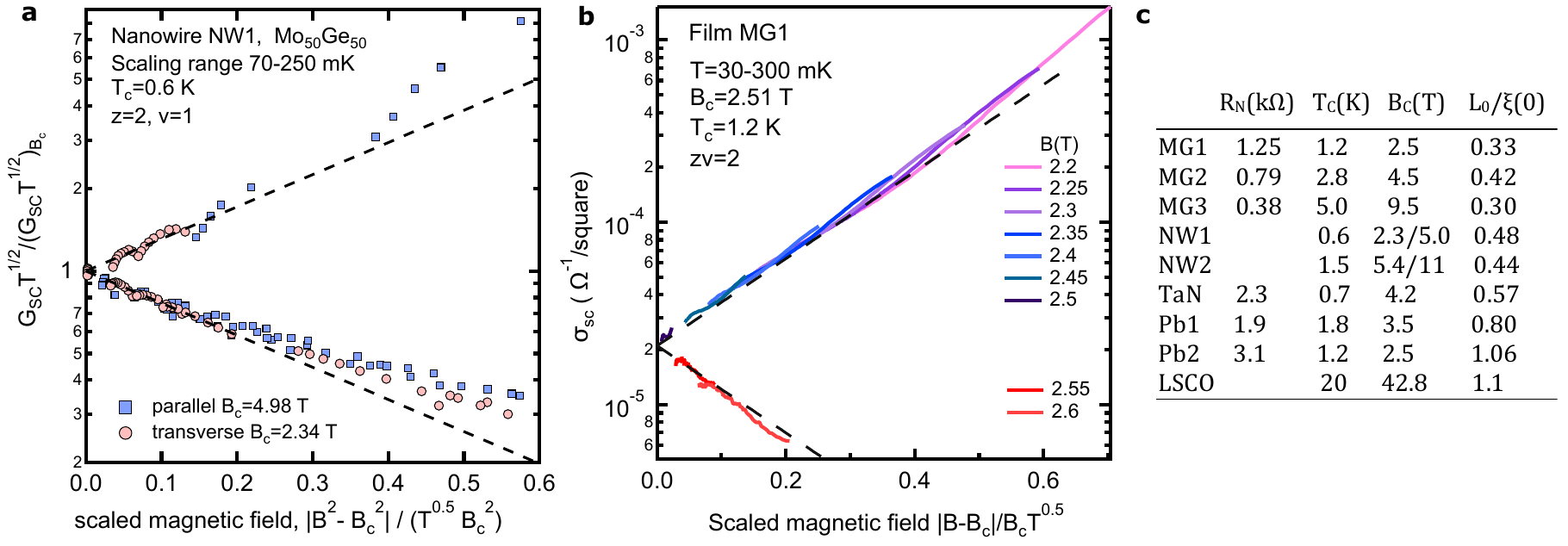}
 \caption{\textbf{Comparison with the theoretical model of the pair-breaking QPT.} 
(a) Scaled and normalized conductance of MoGe nanowire NW1 versus scaled magnetic field.(b) Conductivity of the superconducting channel for MG1 film versus scaled magnetic field. In both panels, the dashed lines have the same absolute slope and are the fits to the theory. (c) Experimental parameters and the extracted ratio  $L_0/\xi\left(0\right)$  for indicated samples. $R_N$ stands for normal state resistance per square. Composition for MoGe samples: Film MG1 Mo$_{50}$Ge$_{50}$, film MG2 Mo$_{78}$Ge$_{22}$, film MG3 Mo$_{78}$Ge$_{22}$, nanowire NW1 Mo$_{50}$Ge$_{50}$, nanowire NW2 Mo$_{78}$Ge$_{22}$. Two numbers of $B_c$ for nanowires indicate fields in perpendicular and parallel orientations.}
 \end{figure*}

We emphasize that $\sigma_{sc}$ represents the superconducting fluctuations only. The subtraction of the dominant contribution of the normal electrons (both the Drude and quantum correction terms) is the critical step in our analysis distinguishing it from all previous work on QPTs in 2D superconducting systems. 

\smallskip
\noindent \textbf{Microscopic scale of the pair-breaking QPT.}

The notable feature of the data shown of Fig. 2(d-e) on the log-linear scale is a “mirror symmetry” between the upper and lower branches. This indicates that the conductivity, when plotted vs $B-B_c$ (rather than $|B-B_c|$) exhibits a smooth exponential variation across $B_c$. 

To explain the observed variation of $\sigma_{sc}$, we have developed a phenomenological model of QPTs based on the conjecture that the scaling theory of localization (STL) \cite{Abrahams_STL} captures the general properties of a one-parameter real-space renormalization group (RG) and, in this regard, can describe QPTs across a wide range of systems. The model, explained in details in the SI and Ref. [\cite{Rogachev_MicroScale}], proposes that the conductivity near a QPT can be approximated as 
\begin{equation}
\sigma=\frac{e^2}{\hbar}(bT^{1/z})^{2-d}g_c\exp{\left( \left(\frac{b}{L_0} \right)^{1/\nu}\frac{B-B_c}{B_c T^{1/z\nu}} \right)}
\end{equation}
Here, $g_c$ is an unknown parameter. The exponent's argument can be written as $(L_{\varphi}(T)/\xi)^{1/\nu}$, where $L_{\varphi}(T)=bT^{-1/z}$ is the dephasing length, and $\xi=L_0(B_c/|B-B_c|)^\nu$ is the correlation length. 

Our analysis goes beyond the standard determination of the critical exponents and focuses on the experimental parameter $(b/L_0)^{1/\nu}$ which determines the slope of the scaled data at $B_c$ on Fig. 3. The parameter  $L_0$, related to the ultraviolet cutoff in the field theories as $\Lambda\approx 1/L_0$, represents the microscopic length where the renormalization flow (integration of the $\beta$-function in STL) starts. A key feature of our model is that $L_0$ is typically known from the microscopic physics of the analyzed system, allowing one to test assumptions about the QPT mechanism and the parameters defining $L_{\varphi}(T)$ by checking whether they reproduce the correct $L_0$.

Originally developed for superconducting films, our model has been validated across a wide range of experimental systems, where we consistently found that the exponential term (or its linear expansion) is a generic term of one-parameter QPTs [\cite{Rogachev_MicroScale}]. With an appropriate choice of $L_{\varphi}(T)$, the extracted $L_0$ values reliably correspond to physically meaningful microscopic scales of QPTs -- such as the mean free path in doped semiconductors and amorphous alloys, the lattice constant in organic semiconductors, the spin spacing in Ising and Heisenberg chains, and the lattice period in optically trapped cold atom systems. The model also captures results from \textit{numerical} studies of the Mott transition via dynamical mean-field theory, treating them as if they were experimental data.

Adapting our model to 2D superconducting films, we note that the theories of the pair-breaking QPT in 1D, 2D and 3D systems yield a dynamical exponent of $z$=2. \cite{RamazashviliColeman_sit,Herbut_z2,Galitski_z2,DelMaestro_AnnPhys} Experimentally we found that $z\nu$$\approx$2, this implies that $\nu$$\approx$1 for films in our films in perpendicular magneit . Since the microscopic pair-breaking processes are similar across dimensions, we adopt the dephasing length expression $L_\varphi\approx(\hbar D/k_BT)^{1/z}$ with $z$=2 as derived by critical pair-breaking theory for nanowires. \cite{DelMaestro_AnnPhys}  Theoretically, the normal state diffusion coefficient $D$ arises from the Cooper pair propagator.  Physically,  $L_\varphi$ represents the distance over which superconducting $\textit{density}$ fluctuation penetrates into the surrounding normal metal. Finally, using standard Drude model and BCS equations (see SI) we obtain: 
\begin{equation}
\sigma_{sc}=\frac{e^2 g_c}{\hbar}{\left(\frac{\hbar D}{k_B T}\right)^{\frac{2-d}{2}}} \exp \left[-1.6\frac{\xi\left(0\right)}{L_0}T_c^{1/2}\frac{B^n-B_c^n}{B_c^nT^{1/2}}\right].
\end{equation}
Here, $\xi(0)$ the zero-temperature Ginzburg-Landau coherence length.
The parameter $n$ comes from the dependence of the pair-breaking strength on magnetic field; $n$=1 for films in a perpendicular field and $n$=2 for nanowires, the data for which have been taken from our recent work \cite{Rogachev_PBnanowire} and included in the analysis. 

 As is required by Eq.3, we use conductivity per square for films (Fig. 3(b) and Fig. 2(d-f)) and conductivity multiplied by $T^{1/2}$ for nanowires (Fig. 3(a)).  The dashed lines, which have the same absolute slope for the upper and lower branches, are the best exponential fit to the data in the proximity of $B_c$. From the slope of the data, and the experimental values of $B_c$ and $T_c$ the ratio $L_0/\xi(0)$ has been obtained. 
 
The second major result of our work is presented in the table in Fig. 3(c), which lists the parameters of the studied systems and presents the experimental $L_0/\xi(0)$ ratio. The ratio is in the range 0.8-1.14 for amorphous Pb films and LSCO and in the range 0.3-0.57 for MoGe films and nanowires and TaN films. The coherence length $\xi(0)$ characterizes the size of a Cooper pair (see our comments on this in the SI). For the pair-breaking QPT where the order parameter starts to grow from zero at $B_c$, $\xi(0)$ is the minimal spatial scale over which superconductivity can exist and, hence, is a natural microscopic length scale. Consistent with this physical expectation, $L_0$ and $\xi(0)$ agree to within a factor of order unity. 

Our analysis can be presented in a more standard way. For the pair-breaking QPT, we expect  $L_0$$\approx$$\xi(0)$ and $L_\varphi=(\hbar D/k_BT)^{1/2}$. Using these values as parameters in our model, we can predict the slope of the scaled experimental data to within a coefficient of order one, with some variation between different materials.

\noindent \textbf{Discussion and conclusions.}

Our analysis leads to the conclusion that the QPTs in the studied films occur via the pair-breaking mechanism. The evidence for this is as follows: 1) The transition takes place in the background of normal electrons which contribute 0.9-0.95 of conductivity at the critical point. 2) The determined dynamical exponent $z$=2 is a signature of a pair-breaking QPT. \cite{RamazashviliColeman_sit,Herbut_z2,Galitski_z2,DelMaestro_AnnPhys} 3) There is overall excellent consistency between QPTs in the films and in the nanowires where a critical pair-breaking microscopic theory is available and matches well to the data. 

We can further conclude that the magnetic-field-driven QPT in La$_{1.92}$Sr$_{0.08}$CuO$_{4}$ also occurs via the pair-breaking mechanism. Hopefully this finding will help to clarify the nature of the superconductivity breakdown in over-doped cuprates in zero field. It is now debated if the breakdown occurs via a ‘dirty’ d-wave BCS pair-breaking mechanism \cite{LembergerRanderia_LSCO_qpt} or via non-BCS mechanism. \cite{Bozovic_LSCO_SF_density}  The former case would be similar to our observations.
 
Leveraging the universality of our model, we have extended it to study QPTs in 1D and 2D Josephson junction (JJ) arrays, which consist of lithographically defined superconducting islands connected by tunneling barriers. We found that the microscopic physics of QPTs in JJ arrays is governed by the picture of the interacting \textit{phase-only} excitations (plasmons) and is fundamentally different from the pair-breaking mechanism [\cite{Rogachev_QPT_JJ}]. In JJ arrays, \textit{entire} sample resistance arises from plasmons. They propagate with velocity $v_{pl}$=$a/\sqrt{L_K C_0}$, where $a$ is the distance between the islands, $L_k$ is the kinetic inductance of a junction, and $C_0$ is the capacitance of an island to the ground. The array's dephasing length is $L_{\varphi}$=$v_{pl}\hbar/k_B T$ and the dynamical critical exponent $z$=1. The experimental microscopic scale was found to match the screening length of a Cooper pair given by $\Lambda$=$a\sqrt{C_1/C_0}$, where $C_1$ is the capacitance between islands.

JJ arrays are inherently granular superconducting systems. Comparing them with the films and nanowires undergoing pair-breaking QPTs, it becomes apparent that the distinguishing and unifying characteristic of the latter group is the absence of weak links that could act as Josephson junctions. All analyzed films were grown under conditions designed to suppress granularity and do not exhibit the so-called giant magnetoresistance peak, a known signature of a non-uniform order parameter.\cite{BaturinaStrunk,OvadiaShahar}

In conclusion, we observed a pair-breaking quantum phase transition in a series of superconducting MoGe, Pb, and TaN films, as well as in single-crystal LSCO. A theoretical model originally developed to describe the quantum critical regime in these systems was also found to capture the microscopic physics of Josephson junction arrays, various complex non-superconducting systems, and even some numerical models of the Mott transition.

A central premise of our model is that, in condensed matter systems, the microscopic scale $L_0$ is physically meaningful and directly influences the slope of the scaled experimental data across a QPT. Thus, our approach is complementary to the standard renormalization group framework in quantum field theories, which aims to find quantities such as correlation functions, masses and critical exponents, without explicit reference to $L_0$ [\cite{Shankar_book}]. (It is worth noting that in elementary particle physics, even the fundamental origin of $L_0$ is unknown and hypothesized to arise from quantum fluctuations in gravity [\cite{Peskin_Schroeder_book}].)  

 Since its development in the 1960s and 1970s, the framework of scale invariance and renormalization has been extended to describe phase transitions in non-equilibrium and dynamical systems, \cite{Non-Eq_PT_book,Heyl_DynamicalQPT,Faggian_PT_LaserFeedBack} quantum entanglement and information flows, \cite{Skinner_QPT_MesInduced,Notarmuzi_PT_Informatio} cortical activity and a range of other phenomena in living organisms.\cite{Munoz_PTliving} We argue that identifying relevant microscopic scales and processes, and incorporating them into the analysis of scaled experimental and numerical data, can uncover behavior beyond the universal patterns described by critical exponents, contributing to a more detailed understanding of critical phenomena in these and many other systems.

 \smallskip
\noindent \textbf{ Acknowledgements.}
A.R. acknowledges Université Grenoble Alpes and Institute Néel, where measurements were performed, for hospitality. The authors thank B. Sacépé for the help with the measurements and extensive feedback on the project and A. Del Maestro for valuable comments. The research was supported by National Science Foundation under awards DMR1904221 and DMR2133014.

\smallskip
\noindent \textbf{Author Contributions.}
AR measured the samples, developed the model, analyzed the data and wrote the manuscript. KD fabricated the samples. KD and SF contributed to data analysis and revisions of the manuscript. We confirm that the manuscript has been read and approved by all named authors and that there
are no other persons who satisfied the criteria for authorship but are not listed.

\smallskip
\noindent \textbf{Competing interests.}
The authors declare no competing interests.

\smallskip
\noindent \textbf{Supplementary information.} Supplementary information is included in this document

\smallskip
\noindent \textbf{Correspondance and requests for materials} should be addressed to Andrey Rogachev.

\noindent \textbf{Supplementary Information.} 
\smallskip

\noindent \textbf{S1. MoGe samples fabrication and measurements.} 

The studied films were made of two a-MoGe alloys with composition Mo$_{78}$Ge$_{22}$ (bulk critical temperature, $T_c\cong7$ K) and Mo$_{50}$Ge$_{50}$ ($T_c\cong3$ K). They were deposited via DC magnetron sputtering using a shadow mask with a width of 500 $\mu$m. Prior to and after the film deposition, a 3-nm-thick layer of a-Ge was deposited without breaking vacuum. 	
Measurements were carried out in a perpendicular magnetic field in a dilution refrigerator equipped with room-temperature feedthrough filters, electrical lines made of lossy miniature stainless steel coaxial cables, low-temperature cooper powder-filters, and capacitance to ground, mounted directly on the sample holder. These measures were taken to prevent exposure of the films to thermal and radiative noise.
\smallskip

\noindent \textbf{S2. Generic scaling equation.}

The scaling theory of localization (STL) was originally developed to describe the conductance in disordered systems of non-interacting electrons at $T=0$.  It introduces the dimensionless conductance, $g$, of a hypercube of size $L^d$, which is related to the conductance, $G$, and conductivity, $\sigma$, as $\sigma=G/L^{d-2}=(e^2/\hbar)(g/L^{d-2})$. The theory makes the conjecture that the conductance of a cube of a bigger size $b^dL^d$ is determined only by $b$ and $g(L)$. In the continuous form, this statement is expressed by the scaling equation for the function $\beta$ where $\beta(g(L))=d\ln {g(L)}/d \ln{L}$. This function describes how the conductance changes or, in the language of the renormalization group, “flows” with increasing system size, starting from some microscopic scale, $L_0$, with conductance, $g_0$. This theory also makes a second conjecture that $\beta(g)$ is monotonic and continuous. 

Our model is based on the conjecture that, near the critical point of a QPT, the STL provides a zero-temperature renormalization group description that is generic for QPTs driven by the evolution of a single parameter (e.g., conductance or coupling constant). Initially proposed in our analysis of QPTs in superconducting films, this conjecture has been validated through analysis the scaled experimental data of more than 30 systems spanning 18 distinct material classes.
   	
Near the critical point of the metal-insulator transition (MIT), that is, near the critical conductance $g_c$,  $\beta$ behaves approximately linearly as $\beta=s\ln{g/g_c}$.  The coefficient, $s$, in this equation is equal to the inverse of the correlation exponent, $\nu$ . Using this linear approximation for $\beta$, we integrate the general scaling equation for $\beta(g(L))$ starting from some microscopic conductance, $g_0$, corresponding to some microscopic length scale, $L_0$. For systems that exhibit an MIT, it is assumed that $L_0\approx\ell$, where $\ell$ is the mean free path.  However, we keep the notation $L_0$ in anticipation that this scale may correspond to different quantities in different systems. We also anticipate that the resulting equation is general and applicable to some 1d and 2d systems, so we keep $d$ as a variable. In the linear regime, the integration goes as 
\begin{align}\label{M3}
& \int_{g_0}^{g(L)}\frac{d\ln{g}}{\beta}\approx\nu\int_{g_0}^{g\left(L\right)}\frac{d\ln{g}}{\ln{g/g_c}}=
\nu\ln{\left(\frac{\ln{g/g_c}}{\ln{g_0/g_c}}\right)}\nonumber\\
& \approx\int_{L_0}^{L}{d\ln{L}}=\ln{\left(\frac{L}{L_0}\right)}
\end{align}
and gives $\ln(g/g_c)=\ln (g_0/g_c)(L/L_0)^{1/\nu}$. Exponentiating both sides and converting the equation to the conductivity of a cube with side $L$, we find that $\sigma=e^2 g_c\exp (\ln (g_0/g_c)(L/L_0)^{1/\nu})/(\hbar L^{d-2})$. 

Then, using the expansion $\ln (g_0/g_c)\approx (g_0-g_c)/g_c$ and an approximation that near the critical conductance, $g_c$, the conductance changes linearly with the driving parameter y as $(g_0-g_c)/g_c\approx (y-y_c)/y_c$
 we obtain 
\begin{equation}\label{M4}
\sigma=\frac{e^2}{\hbar L^{d-2}}g_c\exp{\left(\frac{y-y_c}{y_c}\left(\frac{L}{L_0}\right)^{1/\nu}\right)}
\end{equation}
Equation 4 describes the variation of the conductivity of a system as a function of its size $L$ in the critical regime at zero temperature. 

At finite temperature, thermal fluctuations break the system’s coherence, so the variation given by Eq.4 switches into the Ohmic regime at the dephasing length. Mathematically this means that in Eq. (5), we replace $L$ by $L_{\varphi}(T)$. The dephasing length is generically determined by the temperature and the dynamical exponent, $z$, as $L_\varphi=bT^{1/z}$, where $b$ is depends on system-specific parameters. The theory of QPTs states that the quantum critical regime, which we wish to analyze, is restricted by the condition $L_\varphi<\xi$.  That is why the integration is taken to $L<\xi$ and the linear variation  $\beta=(1/\nu)\ln {(g/g_c)}$ is a good approximation for this regime. With these inputs, the final equation for conductivity in the quantum critical regime is 
\begin{equation}\label{M5}
  \sigma=\frac{e^2}{\hbar}(bT^{1/z})^{2-d}g_c\exp{\left( \left(\frac{b}{L_0} \right)^{1/\nu}\frac{y-y_c}{y_c T^{1/z\nu}} \right)}
\end{equation}
Equation 5 describes both the insulating and metallic regime. The sign in the exponent corresponds to the case when $\left(y-y_c\right)>0$ on the metal side of the transition. 
 
\smallskip

\noindent \textbf{S3. Scaling equation for superconducting films and nanowires.} 

Our analysis is based on the conjecture that Eq. 5 captures the general properties of real-space one-parameter renormalization group and, an in this regard, can be applied to many systems including superconducting films and wires. The dynamic exponent for films is $z$$\approx$2, as suggested by the existing pair-breaking critical theories. The choice of $z$ defines the value of the correlation exponent for films to be $\nu$$\approx$1. 

Microscopically, the pair-breaking processes near QPT are very similar in systems of all dimensions. Capitalizing on this physical expectation, the dephasing length is taken as it was found in the critical pair-breaking theory for nanowires,  $L_\varphi\approx(\hbar D/kT)^{1/2}$. In this theory,
the normal state diffusion coefficient, $D$, comes from the Cooper pair propagator in disordered superconducting systems. 

For a film in the perpendicular magnetic field, the pair-breaking strength is proportional to $B$, so the driving term becomes $(y-y_c)/y_c=(B-B_c)/B_c$,  for nanowires the pair-breaking strength is proportional to  $B^2$ so we have $(y-y_c)/y_c=(B^2-B_c^2)/B_c^2$. For superconducting homogeneous films and nanowires, the only natural choice for $L_0$ is zero-temperature, zero-field Ginzburg-Landau coherence length $\xi(0)$.  This is also the size of a Cooper pair in a disordered superconductor.  From the standard BSC equations, we have $\xi(0)=0.85(\xi_0\ell)^{1/2}$, where $\ell$  is the mean free path, $\xi_0=\hbar v_F/\pi\Delta$ is the Pippard coherence length, $v_F$  the Fermi velocity, and $\Delta$ the superconducting gap related to the critical temperature as $2\Delta=3.53kT_c$. Then, taking the diffusion coefficient as $D=v_F\ell/3$, we find that not-well-known parameters,  $D$, $v_F$ and $\ell$, drop out and general Eq.5 can be expressed for this case as following 
\begin{align}\label{M6}
&\sigma_{sc}={\frac{e^2 g_c}{\hbar}\left(\frac{kT}{\hbar D}\right)}^{\frac{d-2}{2}} \times\nonumber\\ 
&\exp\left[\left(\frac{1}{L_0}\frac{\xi\left(0\right)}{\xi\left(0\right)}\left(\frac{\hbar D}{kT}\right)^{0.5}\right)^{1/\nu}\frac{B_c^n-B^n}{B_c^n}\right]=\nonumber\\
&\frac{e^2 g_c}{\hbar}\left(\frac{kT}{\hbar D}\right)^{\frac{d-2}{2}} \exp\left[\left(-1.6\frac{\xi\left(0\right)}{L_0}T_c^{1/2}\right)^{1/\nu}
\frac{B_c^n-B^n}{B_c^nT^{1/2\nu}}\right]
\end{align}
In this equation $n$=1 stands for films in perpendicular field and $n$=2 for nanowires. A slightly different numerical factor in the exponent was used for Pb films, which comes from the relation $\Delta\approx2.25k_BT_c$ found in [\cite{Dynes}], and for LSCO, where the relation for a  $d$-wave superconductor, $\Delta_{max}\approx2.14k_BT_c$, was used. For these two systems, we find it appropriate to use the 2d expression $D$=$v_Fl/2$.

 \smallskip
\noindent \textbf{S4. Superconductivity, coherence lengths and the size of a Cooper pair.} 

There are several coherence lengths introduced in the theory of superconductivity. The Ginzburg-Landau coherence length is defined as $\xi_{GL}=\xi(0)[(T_c-T)/T_c]^{-1/2}$. It diverges at the critical temperature $T_c$ and characterizes the \textit{recovery length} -- a distance over which a perturbation in the order parameter returns to its uniform value. This length also serves as the correlation length of the superconducting transition. The correlation exponent $\nu=1/2$ reflects the fact that BCS theory is a mean-field theory. The length $\xi(0)$ is typically referred to as the Ginzburg-Landau coherence length at zero temperature. Up to a numerical coefficient of order of unity it corresponds to the size of a Copper pair. This can be understood from the following heuristic arguments. 

Let us first consider the case of a clean (not-disordered) superconductor at zero temperature. Electrons are bound into Cooper pairs with bound energy of order of $\Delta$. Nevertheless, due to a virtual quantum excitation, one electron in a Cooper pair can temporarily become "unbound" and propagate over a time $\hbar/\Delta$, covering a distance $v_F \hbar/\Delta$. This distance provides an estimate of the size of a Cooper pair and, up to a numerical factor of $1/\pi$, corresponds to the Pippard coherence length $\xi_0=\hbar v_F/\pi\Delta$. In the case of the disordered superconductor, the motion of the "unbound" electron becomes diffusive. In this regime, the size of a Cooper pair can be estimates as $(D\hbar /\Delta)^{1/2} \simeq (v_F \hbar \ell /3\Delta)^{1/2} \simeq (\xi_0 \ell)^{1/2}\simeq\xi(0)$.      

\smallskip
\noindent \textbf{S5. Finding the conductivity of the superconducting channel for MoGe films.} 

The task at hand is to find a good approximation for the conductivity of a normal channel, $\sigma_n$. In our studies of nanowires, we have found that at sufficiently high fields, their resistance becomes field independent. As such, we chose the experimental value of $\sigma_{exp}$ at high field to be an approximation for $\sigma_n$. This method is not applicable for films; as one can see from Fig. 1c (main text), the magnetoresistance is not zero at high fields and, moreover, it changes sign from negative at low temperatures to positive at high temperatures.
 
Looking for an approximation for $\sigma_n$ , we have studied a film made of non-superconducting Mo$_{30}$Ge$_{70}$ alloy and tried to fit its magnetoresistance using the theories of quantum corrections. This film displayed an expected logarithmic correction; however, the theory did not provide an accurate quantitative dependence for $R\left(T,B\right)$ curves.  

\begin{figure*}
\centering
 \includegraphics[width= 1.0\textwidth]{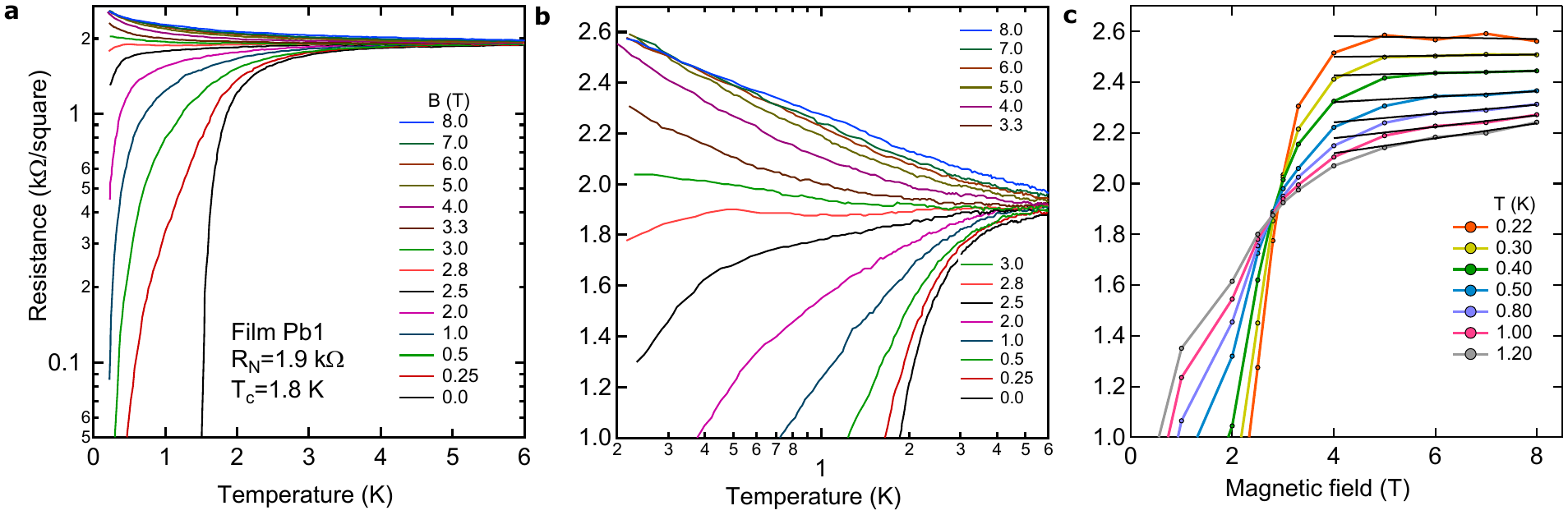}
 \caption{(a) Resistance versus temperature for film Pb1. (b) Same data on log(T) scale to emphasize the transitional regime. The data on both panel have been traced from Fig. 14.12 of Ref. \cite{Xiong_Book} (c) Same data plotted against magnetic field. Please notice that magnetoresistance at high fields is negative at lowest temperature of the measurements.  The solid lines indicate the linear fit to the $R\left(B\right)$ dependence in the range 6-8 Tesla.}
 \end{figure*}

In the end, we developed two empirical methods of extracting $\sigma_{sc}$. In method 1, we employed the formula $\sigma_{sc}(T,B)$=$\sigma_{exp}-\sigma_n$=$1/R(T,B)-1/R(T,B_{max})$, where for $\sigma_n$ we use the inverted experimental resistance at the field that gives the maximum value of $R$ at low temperatures; for example, from Fig.1c (main text) for film MG2, $B_{max}=5$ T. In method 2, we extended the linear fits of the data at high fields to lower fields and obtained the approximation for normal conductance as $\sigma_n$=$1/\left(R\left(T,B_0\right)+A\left(T\right)(B-B_0\right))$; this is shown for the film MG2 as solid lines in Fig.2b (main text).  For MG2, $B_0$=7 T and the coefficient $A\left(T\right)$ was obtained by fitting the data in the interval 6-7 T at several temperatures; a polynomial spline was then used to get the intermediate values of $A\left(T\right)$.  Both methods produced conductivity data with a pronounced insulating branch; in both cases scaling analysis leads to the same critical exponents. Method 1 was used in the initial stage of the project and led to the discovery of the QPT, while method 2 was found to produce more compact data on the scaling plots.  
\smallskip

\noindent \textbf{S6. Conductivity of superconducting fluctuations in amorphous quench-condensed Pb films: extraction from the experimental data and finite-size scaling analysis.} 

\begin{figure*}
\centering
 \includegraphics[width= 1.0\textwidth]{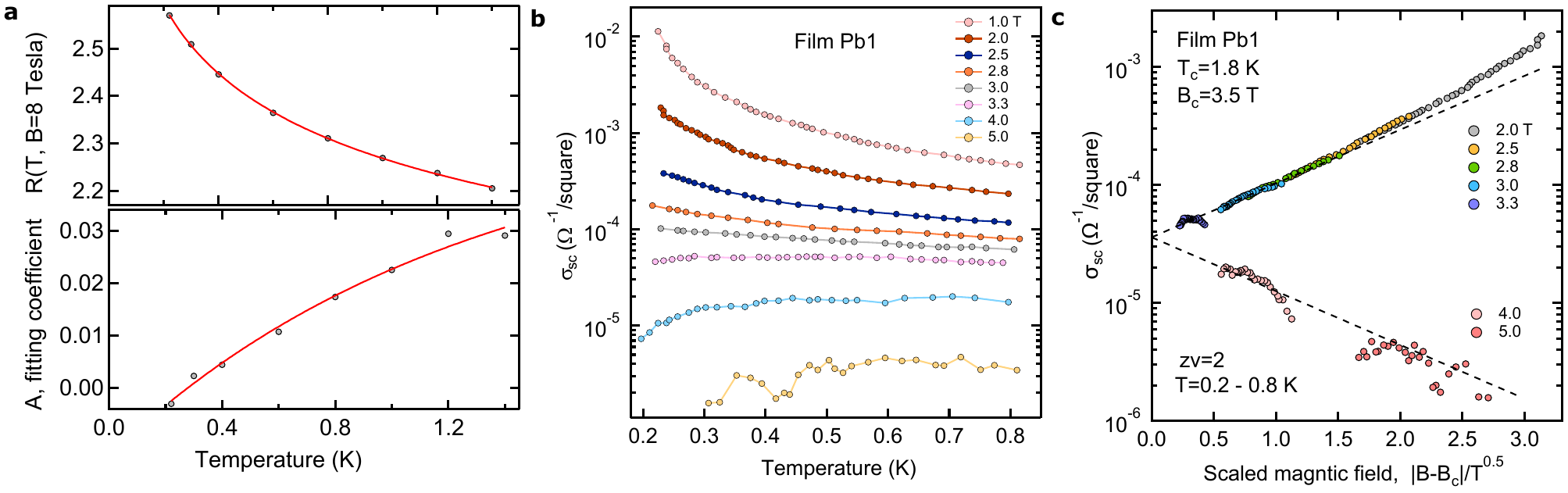}
 \caption{ (a) Interpolation of parameters $R\left(B_0\right)$ and A. (b) Conductivity of the superconducting fluctuations. (c) Scaling plot of $\sigma_{sc}$. The dashed lines have the same absolute slope for superconducting and insulating branches. The numerical value of this slope was used to extract the parameter $\xi(0)/L_0$ as described in the main text.}
 \end{figure*}
 	
We traced the data for two amorphous Pb films published in Ref. [\cite{Xiong_Book}] of the main text. They are re-plotted in Fig. 4. The films were grown inside of a dilution refrigerator by evaporation on substrates kept at low temperatures (4.2-10 K). The resistance of the films was measured with no vacuum breaking and heating them up to higher temperatures. The reference gives the technical details.  The data for the film labeled by us as Pb1 were traced from Fig. 14.12 and for film Pb2 from Fig. 14.11 of this reference. The raw resistance for sample Pb1 needed to be divided by 2 to be consistent with the rest of the samples; probably an incorrect number of squares were used to compute sheet resistance (private communication from P. Xiong). 

In Fig.4c we show the resistance versus magnetic field for film Pb1. As one can notice, the normal state magnetoresistance of the film is negative at $T=0.22$ mK and becomes positive at higher temperatures. This is a common behavior both for MoGe and Pb films.  

In our analysis, we adopt the two-fluid model and assume that in the critical regime the conductivity of the film, $\sigma_{exp}$, is a sum of contributions from the normal and superconducting channels, $\sigma_{exp}$=$\sigma_n+\sigma_{sc}$.  To estimate $\sigma_n$, the following procedure was used. We first fit the magnetoresistance data at high fields to an equation $R(B)$=$R(B_0)+A(B-B_0)$, where $R(B_0)$ and $A$ are fitting parameters. For film Pb1 $B_0$=8 T was used, the fitting was carried out in the range 6-8 T; these linear fits are shown as solid lines in Fig.4c.  The dependence of two fitting coefficients on temperature is shown in Fig.5a; the red lines show interpolation between the points, from which the estimate for normal channel at any given temperature and field can be computed as  $\sigma_n$=$1/[R(T,B_0)+A(T)(B-B_0)]$. The conductivity of the superconducting fluctuations is then computed as $\sigma_{sc}(B,T)$=$\sigma_{exp}(B,T)-\sigma_n(B,T)$ and shown in Fig.5b. Figure 5c shows the scaling plot of $\sigma_{sc}$. 

\end{document}